\documentclass[aps,prd,preprint,noshowkeys,superscriptaddress,nofootinbib]{revtex4-1}

\usepackage{amsmath,amssymb,amsfonts,slashed,cancel}
\usepackage{physics}
\usepackage[dvipdfmx]{graphicx}
\usepackage{mathtools,xcolor}
\usepackage{bm}
\usepackage{tikz}
\usepackage[compat=1.1.0]{tikz-feynhand}

\usepackage[height=8.85in,width=7.0in]{geometry}

\definecolor{BlueViolet}{rgb}{0.2, 0.00, 0.7}
\definecolor{Blue}{rgb}{0.15, 0.00, 0.9}
\usepackage[colorlinks=true,linkcolor=Blue,citecolor=Blue,urlcolor=BlueViolet]{hyperref}

\usepackage[utf8]{inputenc}

\begin{document}

\preprint{YITP-25-69}
\preprint{RIKEN-iTHEMS-Report-25}
\preprint{STUPP-25-278}

\title{An Analytic Prescription for $t$-channel Singularities}

\author{Kento Asai}
\email{kento.asai@yukawa.kyoto-u.ac.jp}
\affiliation{Yukawa Institute for Theoretical Physics, Kyoto University, Kyoto 606--8502, Japan}
\affiliation{Institute for Cosmic Ray Research (ICRR), The University of Tokyo, Kashiwa, Chiba 277--8582, Japan}

\author{Nagisa Hiroshima}
\email{hiroshima-nagisa-hd@ynu.ac.jp}
\affiliation{Department of Physics, Faculty of Engineering Science, Yokohama National University, Yokohama 240--8501, Japan}
\affiliation{RIKEN Center for Interdisciplinary Theoretical and Mathematical Sciences(iTHEMS), RIKEN, Wako 351--0198, Japan}

\author{Joe Sato}
\email{sato-joe-mc@ynu.ac.jp}
\affiliation{Department of Physics, Faculty of Engineering Science, Yokohama National University, Yokohama 240--8501, Japan}

\author{Ryusei Sato}
\email{sato-ryusei-xr@ynu.jp}
\affiliation{Department of Physics, Faculty of Engineering Science, Yokohama National University, Yokohama 240--8501, Japan}
\affiliation{RIKEN Nishina Center, RIKEN, Wako, Saitama 351--0198, Japan}

\author{Masaki J. S. Yang}
\email{mjsyang@mail.saitama-u.ac.jp}
\affiliation{Department of Physics, Faculty of Engineering Science, Yokohama National University, Yokohama 240--8501, Japan}
\affiliation{Department of Physics, Faculty of Science, Saitama University, Saitama 338--8570, Japan}

\begin{abstract}
The $t$-channel singularity is a divergence in the scattering amplitude which occurs when a stable particle propagating in $t$-channel scattering process becomes an on-shell state. 
Such situations appear either in the system of collider experiments or in the context of the cosmological particle production. 
No scheme which is generally applicable is known. 
In this work, we propose a new formulation to identify and remove the source of the divergence. 
The scheme is fully analytical and various applications can be expected. 
This work provides a valuable tool in this research field.
\end{abstract}

\date{\today}

\maketitle

\section{Introduction}
\label{sec:intro}

An on-shell propagator in a scattering process often appears in a variety of systems~\cite{Lehmann:1954rq, Cutkosky:1960sp}. 
The propagator makes the scattering amplitude divergent since it picks a pole. 
Depending on the diagrams of the process, we see two types of singularities: the $s$-channel type and the $t$-channel type~\cite{Mandelstam:1958xc}. 
The singularities in the $s$-channel process~\cite{Breit:1936zzb} have been widely discussed in the context of the 
baryogenesis~\cite{Yoshimura:1978ex} and/or leptogenesis~\cite{Fukugita:1986hr}. A key process for leptogenesis is the lepton-number violating scattering.
During the processes, right-handed neutrinos can appear as on-shell propagators in the intermediate state. 
In most setups, the finite decay width of right-handed neutrinos cures the divergence 
by the so-called narrow-width approximation~\cite{Kolb:1979qa,Giudice:2003jh}.
The other type of the singularity, the $t$-channel one, was first pointed out in the context of hadronic processes of the reaction $\pi N^{*} \to \pi N^{*}$~\cite{Peierls:1961zz}. 
In Ref.~\cite{Brayshaw:1978xt}, it was shown that such a singular behavior is general when the mediator satisfies the on-shell condition and the process can be decomposed into two sub-processes with real particles. 
For example, the process $\mu^{+} \mu^{-} \to W^{+} e^{-} \bar \nu_{e}$ in muon colliders is regarded as such a situation since the propagating $\nu_{\mu}$ can be a real particle and the process is decomposed into $\mu^{-} \to \nu_{\mu} e^{-} \bar \nu_{e}$ and $\nu_{\mu} \mu^{+} \to W^{+}$.
Several methods for regularizing this divergence have been discussed in the literature~\cite{ Melnikov:1996na, Melnikov:1996iu, Melnikov:1996ft, Dams:2002uy, Dams:2003gn,Nowakowski:1993iu,Karamitros:2022nnh}. 
In the context of collider experiments, it is argued that the amplitudes are effectively regularized because the beam sizes are finite. 

For $s$-channel processes, the decay and the inverse decay process are resummed into the propagator as
${1/(p^2-m^2+im\Gamma)}$, 
where $\Gamma$ is the decay width of the $s$-channel particle, and consequently the singularity is always regulated.
In contrast, for $t$-channel processes, the intermediate state is not necessarily an unstable particle. 
If the particle can decay, a similar regularization by its decay width is possible. 
However, for stable particles such as neutrinos, which do not decay, this regularization is not applicable. 
Even when loop corrections are taken into account, a stable particle such as the neutrino cannot acquire a finite decay width. 
This implies that the issue under discussion is not restricted to leading-order perturbative contributions.

In cosmological contexts, more careful treatments of the divergence are needed. 
When multiple species of unstable particles can share the same daughter particles,
the effects of the $t$-channel singularity are more significant. 
However, the concept of the finite beam-width cannot be applied in cosmological systems. 
Instead, 
the interaction of the propagating particle with background gas generates an imaginary contribution to the self-energy and plays a role as the decay width of the quasiparticle~\cite{Grzadkowski:2021kgi,Iglicki:2022jjf}. 
Another idea is to incorporate the decay width into the momentum of the initial state particle. 
It is understood as an imaginary part appearing in the propagator~\cite{Ginzburg:1995bc}. 
Or, the scattering is sometimes just neglected as a subdominant contribution compared to the processes of decays and inverse decays~\cite{Escudero:2019gzq, Escudero:2018mvt}.

In our previous work~\cite{Asai:2023ajh}, we proposed a method to address the issue by using the Feynman's prescription. 
Since the on-shell propagator appears as the divergence proportional to $1/\epsilon$, by evaluating the $1/\epsilon$ term and removing it, we can separate scattering production of the particles from those by decays and inverse decays. 
It was practically achieved by performing numerical computations. In this paper, we show that the separation can be done in an analytic way. 
Our new scheme significantly reduces the computational cost and enables us to investigate a huge parameter space of cosmological systems with $t$-channel singularities. 

The structure of this paper is as follows. 
The next section describes the $t$-channel singularity and our prescription, and in Sec.~\ref{sec:application} , we apply it to the U(1)$_{L_\mu-L_\tau}\times$U(1)$_L$ model and calculate the cosmological production of Majoron. We summarize in Sec.~\ref{s:discussion}.

\section{Analytic calculation of $t$-channel singularity}
\label{sec:tchannel}

\subsection{Field theoretical  interpretation of the $t$-channel singularity}
\label{sec:tchannelexample}

First, we comment on the physical interpretations of the $t$-channel singularity in the context of the quantum field theory. 
When the propagator in a $t$-channel scattering process goes on-shell, the process effectively corresponds to a sequence of an inverse decay followed by a decay. 
According to the standard interpretation, this divergence originates from an on-shell particle that propagates indefinitely. 

\begin{figure}[tbp]
    \centering
\begin{tikzpicture}
\begin{feynhand}
	\vertex [particle] (fig2_i1) at (1, 1.5) {$\phi_3$};
	\vertex [particle] (fig2_i2) at (1, -1.5) {$\phi_4$};
	\vertex [particle] (fig2_f1) at (4.5, -1.5) {$\phi_1$};
	\vertex [particle] (fig2_f2) at (4.5, 1.5) {$\phi_2$};
	\vertex (fig2_g1) at (2.5, 0.5) ;
	\vertex (fig2_g2) at (3, -0.5) ;
    \vertex (constant_1) at (2.5, 0.8) {$y_1$};
	\vertex (constant_2) at (3, -0.8) {$y_2$};
    \vertex (constant_2) at (2.5, -0.1) {$p$};
	\propag [scalar] (fig2_i1) to [edge label = \text{$p_3$}] (fig2_g1);
	\propag [scalar] (fig2_i2) to [edge label = \text{$p_4$}] (fig2_g2);
	\propag [scalar] (fig2_g1) to [edge label = \text{$\Phi$}] (fig2_g2);
	\propag [scalar] (fig2_g2) to [edge label = \text{$p_1$}] (fig2_f1);
	\propag [scalar] (fig2_g1) to [edge label = \text{$p_2$}] (fig2_f2);
\end{feynhand}
\end{tikzpicture}
\caption{
A $t$-channel diagram for scattering process $\phi_3\phi_4\to \phi_1\phi_2$ for scalar fields $\phi_i$ with momenta $p_i ~ (i=1\sim4)$. 
There is a stable scalar particle $\Phi$ with momentum $p$ in the intermediate state.
The coupling among $\phi_2, \phi_3,$ and $\Phi$ ($\phi_1, \phi_4$, and $\Phi$) is represented by $y_1 (y_2)$.
}
\label{fig:scalar}
\end{figure}
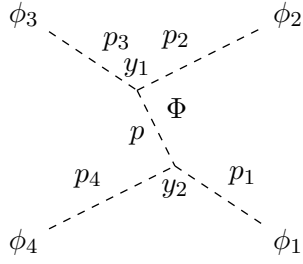

We showcase the structure of the divergence by taking a simple example of Fig.~\ref{fig:scalar}.
There is an intermediate state $\Phi$ with mass $m$ and other scalars $\phi_i$ with masses $m_i$ and momenta $p_i ~ (i=1\sim4)$.
The coupling among $\phi_2, \phi_3,$ and $\Phi$ ($\phi_1, \phi_4$, and $\Phi$) is represented by $y_1 (y_2)$.
When the conditions that $m_3>m+m_2$ and $m_1>m+m_4$ are satisfied, the decay
($\phi_3\to\Phi+\phi_2$) and the inverse decay ($\Phi+\phi_4\to \phi_1$) can occur.

From the Fig.~\ref{fig:scalar}, the
scattering cross section of the process $\phi_3\phi_4\to \phi_1\phi_2$ is given
by 
\begin{align}
\sigma &= 
\dfrac{1}{2E_32E_4}\int \dd\Pi_{1} \dd\Pi_{2}(2 \pi)^{4} \delta^{(4)}\left(p_{1}+p_{2}-p_{3}-p_{4}\right) \sum\left|\mathcal{M}_{\phi_3\phi_4\to \phi_1\phi_2}\right|^{2} 
\nonumber \\
&=
\dfrac{1}{2E_32E_4}\int \dd\Pi_{1} \dd\Pi_{2}(2 \pi)^{4} \delta^{(4)}\left(p_{1}+p_{2}-p_{3}-p_{4}\right) 
y_1^2\left| \dfrac{i}{p^2-m^2+i\varepsilon} \right|^2y_2^2\, ,
\end{align}
where $\dd\Pi_{i} =\frac{1}{(2 \pi)^{3}} \frac{\dd^{3} \bm{p}_{i}}{2 E_{i}}$ denotes the phase space integral.
The momentum-energy conservation at each vertex leads to, 
\begin{align}
\sigma &=
\dfrac{y_1^2 y_2^2}{2E_32E_4}\int \dd\Pi_{1} \dd\Pi_{2}(2 \pi)^{4} \dd^4p \,
\delta^{(4)}\left(p_{3}-p-p_{2}\right) 
\delta^{(4)}\left(p+p_{4}-p_{1}\right) 
\left| \dfrac{i}{p^2-m^2+i\varepsilon} \right|^2
\nonumber \\
&= \dfrac{y_1^2 y_2^2}{2E_32E_4} \int \dd\Pi_{1} \dd\Pi_{2}(2 \pi)^{4} \, \textrm{p.v.}\Bigm\vert_{p^2\neq m^2} \left( \int \dd^4p \, \delta^{(4)}\left(p_{3}-p-p_{2}\right) \delta^{(4)}\left(p+p_{4}-p_{1}\right) \frac{1}{(p^2 - m^2)^2} \right) 
\nonumber \\
&\quad + \dfrac{y_1^2 y_2^2}{2E_32E_4} \lim_{\eta \to 0^+} \int \dd\Pi_{1} \dd\Pi_{2}(2 \pi)^{4} \int_{m^2-\eta < p^2 < m^2+\eta} \hspace{-10mm} \dd^4p \, \delta^{(4)}\left(p_{3}-p-p_{2}\right) \delta^{(4)}\left(p+p_{4}-p_{1}\right) \dfrac{1}{(p^2-m^2)^2+\varepsilon^2}
\nonumber \\
&\equiv \sigma_{\textrm{off-shell}} + \sigma_{\textrm{on-shell}}~,
\end{align}
where p.v.$\bigm\vert_{p^2\neq m^2}$ denotes Cauchy's principal value defined by
\begin{equation}
   \textrm{p.v.}\Bigm\vert_{p^2\neq m^2} \left( \int \dd^4p \, f(p) \right) 
   \equiv \lim_{\eta \to 0^+} \left( \int_{p^2 \leq m^2 - \eta} \dd^4 p + \int_{p^2 \geq m^2 + \eta} \dd^4 p \right) f(p)~,
\end{equation}
for $\eta \gg \varepsilon$. 
Taking the limit of $\varepsilon \to 0$ before that of $\eta \to 0$, the on-shell part $(p^2=m^2)$ which corresponds to the divergent term is
\begin{align}
    \sigma_{\textrm{on-shell}} 
    &=
    \frac{y_1^2 y_2^2}{2E_32E_4} \int \dd\Pi_{1} \dd\Pi_{2}(2 \pi)^{4} 
    \int_{m^2-\eta < p^2 < m^2+\eta} \hspace{-10mm} \dd^4p \, \delta^{(4)}\left(p_{3}-p-p_{2}\right) \delta^{(4)}\left(p+p_{4}-p_{1}\right) \frac{\pi}{\varepsilon} \frac{1}{\pi} \frac{\varepsilon}{(p^2-m^2)^2+\varepsilon^2}
    \nonumber \\
    & =
    \frac{y_1^2 y_2^2}{2E_32E_4} \int \dd\Pi_{1} \dd\Pi_{2}(2 \pi)^{4} 
    \int_{m^2-\eta < p^2 < m^2+\eta} \hspace{-10mm} \dd^4p \, \delta^{(4)}\left(p_{3}-p-p_{2}\right) \delta^{(4)}\left(p+p_{4}-p_{1}\right)  \dfrac{\pi}{\varepsilon}\delta (p^2-m^2)
    \nonumber \\
    & =
    \frac{\pi}{\varepsilon} \frac{y_1^2 y_2^2}{2E_32E_4}
    \int \dd\Pi_1 \dd\Pi_2 \frac{\dd^3 \bm{p}}{2E} \, (2 \pi)^{4} \delta^{(4)}\left(p_{3}-p-p_{2}\right) \delta^{(4)}\left(p+p_{4}-p_{1}\right) \, .
\end{align}
In the above calculation, we use the following formulas:
\begin{equation}
   \lim_{\varepsilon \to 0^+} \frac{1}{\pi} \frac{\varepsilon}{(p^2 - m^2)^2 + \varepsilon^2}
   = \delta(p^2 - m^2)~,\ \textrm{and}\ \dd^4p  \ \delta(p^2-m^2)=\frac{\dd^3 \bm{p}}{2E}~.
\end{equation}
By exchanging the order of the integration, we obtain 
\begin{align}
\label{eq:xsec_on-shell}
   \sigma_{\textrm{on-shell}}
   &=
   \frac{1}{\varepsilon} E \cdot \frac{y_1^2}{2E_3} \int \dd\Pi_2 \dfrac{\dd^3 \bm{p}}{2E(2\pi)^3}
   (2\pi)^4
   \delta^{(4)}\left(p_{3}-p-p_{2}\right) 
   \cdot
    \frac{y_2^2}{2E_42E}
    \int \dd\Pi_1
    (2\pi)^4
    \delta^{(4)}\left(p+p_{4}-p_{1}\right)
    \nonumber \\
    &=
    \frac{1}{\varepsilon} E \cdot
    \Gamma(\phi_3\to \Phi\phi_1) \cdot
    \sigma(\phi_4\Phi\to \phi_2) \, .
\end{align}
As explicitly shown in Eq.~\eqref{eq:xsec_on-shell}, the divergence with $1/\varepsilon$ dependence corresponds to the on-shell contribution, that is, the sequence of a decay and an inverse decay.

\subsection{Prescriptions in the literature}
\label{subsec:prescription}

\begin{figure}[tbp]
\begin{center}
\begin{tikzpicture}
\begin{feynhand}
    \vertex [particle] (fig1_i1) at (1   ,  1.5);
	\vertex [particle] (fig1_i2) at (1   , -1.5);
	\vertex [particle] (fig1_f1) at (4.5, -1.5);
	\vertex [particle] (fig1_f2) at (4.5,  1.5);
	\vertex (fig1_g1) at (2.5,  0.5);
	\vertex (fig1_g2) at (3.  , -0.5);
	\propag [plain] (fig1_i1) to [edge label = \text{$p_3$}] (fig1_g1);
	\propag [plain] (fig1_i2) to [edge label = \text{$p_4$}] (fig1_g2);
	\propag [plain] (fig1_g1) to (fig1_g2);
	\propag [plain] (fig1_g2) to [edge label = \text{$p_1$}] (fig1_f1);
	\propag [plain] (fig1_g1) to [edge label = \text{$p_2$}] (fig1_f2);
\coordinate (A) at (5.5,0); 
\draw (A) node[left, font = \huge] {=}; 
\draw [->](2,-2.5)to (3.5,-2.5)node[right]{$t$};;

    \vertex [particle] (fig1_i1_2) at (6   , 2);
	\vertex [particle] (fig1_f2_2) at (9.5,  2);
	\vertex (fig1_g1_2) at (7.5,  1);
	\vertex (fig1_g2_2) at (9.  , -0.5 );
	\propag [plain] (fig1_i1_2) to [edge label = \text{$p_3$}] (fig1_g1_2);
	\propag [plain] (fig1_g1_2) to [edge label = \text{$p_3-p_2$}](fig1_g2_2);
	\propag [plain] (fig1_g1_2) to [edge label = \text{$p_2$}] (fig1_f2_2);
\coordinate (B) at (10.2,0); %
\draw (B) node[left, font = \huge] {+}; 

	\vertex [particle] (fig1_i2_3) at (10.5   , -2);
	\vertex [particle] (fig1_f1_3) at (14, -2);
	\vertex (fig1_g1_3) at (10.5,  0.5);
	\vertex (fig1_g2_3) at (12.5  , -1);
	\propag [plain] (fig1_i2_3) to [edge label = \text{$p_4$}] (fig1_g2_3);
	\propag [plain] (fig1_g1_3) to [edge label = \text{$p_1-p_4$}] (fig1_g2_3);
	\propag [plain] (fig1_g2_3) to [edge label = \text{$p_1$}] (fig1_f1_3);
\draw [->](9,-2.5)to (10.5,-2.5)node[right]{$t$};
\end{feynhand}
\end{tikzpicture}
\end{center}
\caption{
Decomposition of the $t$-channel scattering process. The left-hand side is regarded as a sum of the two diagrams in the right-hand side when the intermediate state becomes on shell. The first(second) diagram in the right-handed side corresponds to the decay(inverse-decay). We denote the momenta of incoming and outgoing particles by $(p_3,p_4)$ and $(p_1,p_2)$, respectively.}
\label{fig:diagram-decomposition}
\end{figure}
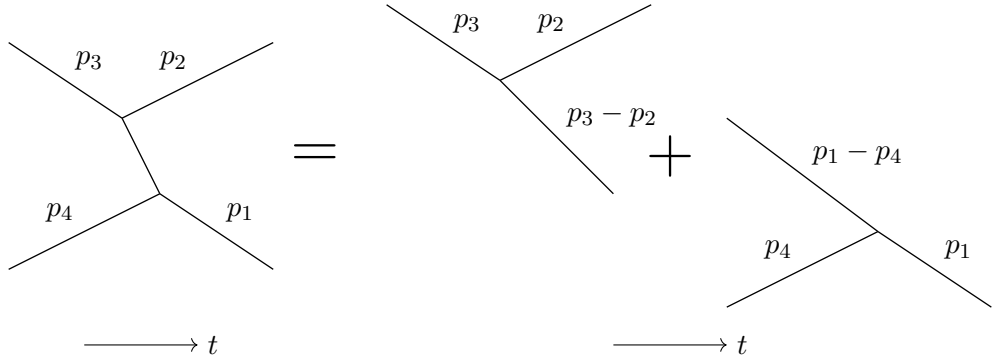

Before explaining our scheme to eliminate the $t$-channel singularity, we briefly review the general structure of this singularity below. 
The scattering amplitude of a $t$-channel diagram contains a propagator that can be on shell. 
A general expression of the scattering amplitude is
\begin{equation}
\label{eq:amp_general}
	\left|\mathcal{M}^t\right|^2 = \frac{1}{ ((p_3 -p_2)^2-m^2)^2+\epsilon^2}f(p_1,p_2,p_3,p_4) \, , 
\end{equation}
where the assignments of the momenta of the initial and final states are shown in Fig.~\ref{fig:diagram-decomposition}.
As it is apparent in Eq.~\eqref{eq:amp_general}, the amplitude diverges when the mediator becomes on-shell. 
To analyze its mathematical structure, we introduce a new variable $X$ as 
\begin{equation}
    X= (p_3 -p_2)^2-m^2 \, . 
\end{equation}
Physical quantities, such as the scattering cross section or the decay width, are obtained by performing the integral of the squared amplitude
\begin{equation}
\label{eq:int_t-amplitude}
    \int_a^b \dd X \left|\mathcal{M}^t\right|^2 = 
    \int_a^b \dd X \frac{1}{ X^2+\epsilon^2}f(p_1,p_2,p_3,p_4)\, ,
\end{equation}
where we denote the lower (upper) limit of the integral with $a (b)$. 
Note that $X$ can be negative since it is an inner product of a four-momentum. The singularity appears when the integral is performed in a range $a<0<b$. 

The regularization of the amplitude is especially difficult when the intermediate state is stable and on-shell. 
Several prescriptions are proposed in the literature. 
For example:  

\begin{itemize}
    \item By considering a system like collider experiments, the finite beam-size effect removes the singularity~\cite{Melnikov:1996iu,Melnikov:1996na}. 
    However, it is not applicable to cosmological systems where the concepts of the beam width cannot be introduced.  
    
    \item The singularity can be removed if the mass of the initial unstable particle appears as the imaginary part, which makes a shift of $m\to m- i\Gamma$~\cite{Ginzburg:1995bc}. 
    However, this violates the energy-momentum conservation at vertices. 
    
    \item If the particle of interest propagates interacting with background particles, the mean free path of the propagating particle induces a finite imaginary width~\cite{Grzadkowski:2021kgi,Iglicki:2022jjf}. 
    It can be applicable only at high-energy regimes since this corresponds to the regularization by the finite-temperature effect.
\end{itemize}

There are no schemes suitable to investigate the particle interactions in a low-temperature Universe. 
Furthermore, we have to be careful about the double-counting. 
In the Boltzmann equation describing the particle production, the scattering incorporates decays and inverse decays of particles, meaning that the contribution needs to be subtracted when we evaluate the yields of particles.

In Ref.~\cite{Asai:2023ajh}, the issue is addressed by adopting the Feynman prescription. 
The Feynman prescription regards the divergence as the propagation of a real particle. 
It is graphically understood as Fig.~\ref{fig:diagram-decomposition}. 
The scattering diagram (left) is decomposed into two pieces of decay and inverse decay (right)
as shown in Sec.~\ref{sec:tchannelexample} with a toy model. 
By treating the diagrams independently, the double-counting of the processes can be avoided. 
Then by specifying the part proportional to $1/\epsilon$, which is responsible for decays and inverse decays, and by removing it, the pure contribution from the scattering on particle production is obtained as the remaining. 
The evaluation of the $1/\epsilon$ term was numerically done in Ref.~\cite{Asai:2023ajh}.

\subsection{Analytic prescription for $t$-channel singularity}
\label{subsec:analytic-prescription}
In our analytic scheme, we redefine the variable $X$ as $X=\varepsilon \bar{X}$. 
Then the integral of Eq.~\eqref{eq:int_t-amplitude} becomes 
\begin{equation}
	\int_{a/\varepsilon}^{b/\varepsilon} \dd(\varepsilon \bar{X} ) \frac{1}{ (\epsilon \bar{X})^2+\epsilon^2}g(p_1,p_2,p_3,p_4)
	=\int_{a/\varepsilon}^{b/\varepsilon} \dd(\varepsilon \bar{X} )\dfrac{1}{\epsilon ^2}\frac{1}{ \bar{X}^2+1}g(p_1,p_2,p_3,p_4) \, .
\end{equation}
The function $g$, as well as $\bar{X}$, is expressed in terms of inner products of four-momenta since these quantities are Lorentz invariant. 
As a consequence, the function $g$ is expressed as powers of $\bar{X}$. 
Extracting the zeroth-order term of $\bar{X}$ in the function $g$, which is the source of the divergence in the $t$-channel singularity, we obtain
\begin{equation}
	\int_{a/\varepsilon}^{b/\varepsilon} \dd (\epsilon \bar{X}) \dfrac{1}{\epsilon ^2}\frac{1}{ \bar{X}^2+1}
    =
	\dfrac{1}{\epsilon }\left(\arctan\left(\dfrac{b}{ \epsilon}\right)-\arctan\left(\dfrac{a}{ \epsilon}\right)\right) \, . \label{after_int}
\end{equation}
By performing an asymptotic expansion of the function following a standard manner, we obtain
\begin{equation}
\arctan\dfrac{x}{\epsilon} = \text{sgn}(x)\dfrac{\pi}{2} - \arctan\dfrac{\epsilon}{x} = \text{sgn}(x)\ \dfrac{\pi}{2} - \dfrac{\epsilon}{x} + \mathcal{O} (\epsilon^3) \, ,
\end{equation}
where $\text{sgn}(x)$ returns the sign of $x$. Further manipulation is done case by case, depending on the sign of the product $ab$. 

\begin{enumerate}
\item $ab > 0$ 
\begin{equation}
	\dfrac{1}{\epsilon }\left(\pm  \dfrac{\pi}{2} - \dfrac{\epsilon}{b} + \mathcal{O} (\epsilon^3)-\left(\pm  \dfrac{\pi}{2} - \dfrac{\epsilon}{a} + \mathcal{O} (\epsilon^3)\right)\right)
	=\dfrac{1}{a} - \dfrac{1}{b} +  \mathcal{O} (\epsilon^2)\, ,
\end{equation}
where a double sign is in the same order. No divergence appears in this case. 

\item $ab < 0$
\begin{equation}
	\dfrac{1}{\epsilon }\left(- \dfrac{\pi}{2} - \dfrac{\epsilon}{b} + \mathcal{O} (\epsilon^3)-\left( \dfrac{\pi}{2} - \dfrac{\epsilon}{a} + \mathcal{O} (\epsilon^3)\right)\right)
	=-  \dfrac{\pi}{\epsilon } + \dfrac{1}{a} -\dfrac{1}{b} + \mathcal{O} (\epsilon^2)\, .
\end{equation}
The effect from the on-shell mediator, $X=0$, appears in the first term of whose contribution is counted in the decay and inverse decay calculations. 
Hence by removing the first term, we obtain 
\begin{equation}
	\dfrac{1}{a} -\dfrac{1}{b} + \mathcal{O} (\epsilon^2)\, .
\end{equation}
\item $a b = 0$

In this case, $\epsilon$ is a finite but nonzero quantity. 
The limit $\epsilon \to 0$ should be taken after the integration:

\begin{equation}
\tan^{-1}\dfrac{0}{\epsilon} = \tan^{-1}0 = 0\, .
\end{equation}

For instance, if $b=0$, we see 
\begin{equation}
	\dfrac{1}{\epsilon }\left(0 - \left( \dfrac{\pi}{2} - \dfrac{\epsilon}{a} + \mathcal{O} (\epsilon^3)\right)\right)
	=-  \dfrac{\pi}{2\epsilon } + \dfrac{1}{a} + \mathcal{O} (\epsilon^2) \, .
\end{equation}
The first term corresponds to that from $X=0$, i.e., the on-shell mediator. 
It should be removed as we have explained in the previous case. 
The quantity we need is  
\begin{equation}
	 \dfrac{1}{a} + \mathcal{O} (\epsilon^2)\, .
\end{equation}
\end{enumerate}
Hence, for each case, we obtain an analytical prescription to remove the $t$-channel singularity in the scattering cross section.

\section{Application to U(1)$_{L_\mu-L_\tau}\times$U(1)$_L$ model}
\label{sec:application}
Our scheme can be applied to models beyond the Standard Model (SM). Here we demonstrate its applicability by taking a model with additional U(1)$_{L_\mu-L_\tau}\times$U(1)$_L$ symmetry~\cite{Araki:2019rmw}. 
In the U(1)$_{L_\mu-L_\tau}\times$U(1)$_L$ symmetric model, both the U(1)$_{L_\mu-L_\tau}$ gauge boson, $Z'$, and a Majoron $\phi$ are unstable and decay into neutrinos, which are stable particles.
For the calculation of the contribution to the Majoron production through the Comptonlike scattering as $Z' \nu \to \phi \bar{\nu}$, the intermediate neutrino propagating in the $t$-channel diagram becomes on-shell at a certain momentum.
The production cross section diverges at that momentum, and the remedy discussed in Sec.~\ref{sec:tchannel} should be implemented.

In this class of model, a $Z'$-boson corresponding to the U(1)$_{L_\mu-L_\tau}$ gauge symmetry and a Majoron $\phi$ as a pseudo-Nambu-Goldstone boson of the global lepton number U(1)$_L$ symmetry breaking are introduced. 
Denoting the SM Lagrangian with ${\cal L}_{\rm SM}$, 
the Lagrangian of the model is given as~\cite{Araki:2021xdk,Asai:2023ajh}
\begin{align}
   \mathcal{L} &= \mathcal{L}_{\rm SM} + \mathcal{L}_{Z'} + \mathcal{L}_\phi~, \\
\label{eq:Lag-Zmt}
    \mathcal{L}_{Z'}
    &= -\frac{1}{4} Z'^{\rho\sigma} Z'_{\rho\sigma} + \frac{1}{2} m_{Z'}^2 Z'^\rho Z'_\rho + g_{Z'}^{} Z'^\rho J_{Z'}^\rho + \epsilon e {Z'}_\mu J_{\rm EM}^\mu~, \\
\label{eq:Lag-phi}
    \mathcal{L}_\phi 
    &= -\frac{1}{2}m _\phi^2 \phi^2 + (h_{\alpha\beta} \bar{\nu}_\alpha \nu^c_\beta \phi + \mathrm{H.c.})~, 
\end{align}
where $m_{Z'}$ and $m_\phi$ is mass of the $Z'$ and $\phi$, respectively, and $Z'^{\rho\sigma}$ is the field strength tensor of $Z'_\mu$. 
Introducing the charge conjugate matrix $C$, the charge conjugation of neutrinos is expressed as $\nu_\alpha^c \equiv ( \nu_\alpha^{})^c = C\bar{\nu}_\alpha^T$. 
The expression of the electromagnetic and U(1)$_{L_\mu-L_\tau}$ currents are 
\begin{align}
\label{eq:EM-current}
    J_{\rm EM}^\rho &= 
    \sum_{i=1,2,3} \left( \frac{2}{3} \bar{u}_i \gamma^\rho u_i - \frac{1}{3} \bar{d}_i \gamma^\rho d_i - \bar{e}_i \gamma^\rho e_i \right)~, \\
    J_{Z'}^\rho &= 
    \bar{\mu} \gamma^\rho \mu + \bar{\nu}_\mu \gamma^\rho P_L \nu_\mu - \bar{\tau} \gamma^\rho \tau - \bar{\nu}_\tau \gamma^\rho P_L \nu_\tau~,
\end{align}
where $i$ corresponds to the index of the fermion generation.

\subsection{Invariant matrix elements with $t$-channel singularity}
\label{subsec:matrix-element}
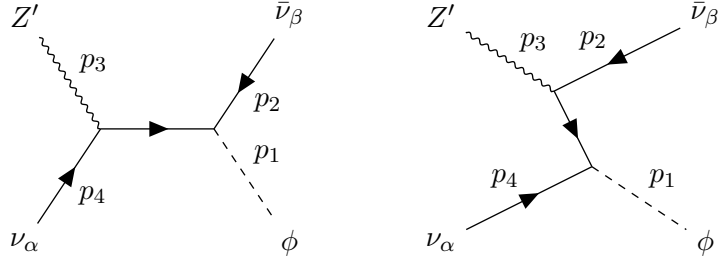
\begin{figure}[tbp]
\begin{center}
\begin{tikzpicture}
\begin{feynhand}
	\vertex [particle] (fig1_i1) at (-4.5, 1.5) {$Z'$};
	\vertex [particle] (fig1_i2) at (-4.5, -1.5) {$\nu_\alpha$};
	\vertex [particle] (fig1_f1) at (-1, -1.5) {$\phi$};
	\vertex [particle] (fig1_f2) at (-1, 1.5) {$\bar{\nu}_\beta$};
	\vertex (fig1_g1) at (-3.5, 0);
	\vertex (fig1_g2) at (-2, 0);
	\propag [photon] (fig1_i1) to [edge label = \text{$p_3$}] (fig1_g1);
	\propag [fermion] (fig1_i2) to [edge label' = \text{$p_4$}] (fig1_g1);
	\propag [fermion] (fig1_g1) to (fig1_g2);
	\propag [scalar] (fig1_g2) to [edge label = \text{$p_1$}] (fig1_f1);
	\propag [anti fermion] (fig1_g2) to [edge label' = \text{$p_2$}] (fig1_f2);

	\vertex [particle] (fig2_i1) at (1, 1.5) {$Z'$};
	\vertex [particle] (fig2_i2) at (1, -1.5) {$\nu_\alpha$};
	\vertex [particle] (fig2_f1) at (4.5, -1.5) {$\phi$};
	\vertex [particle] (fig2_f2) at (4.5, 1.5) {$\bar{\nu}_\beta$};
	\vertex (fig2_g1) at (2.5, 0.5);
	\vertex (fig2_g2) at (3, -0.5);
	\propag [photon] (fig2_i1) to [edge label = \text{$p_3$}] (fig2_g1);
	\propag [fermion] (fig2_i2) to [edge label = \text{$p_4$}] (fig2_g2);
	\propag [fermion] (fig2_g1) to (fig2_g2);
	\propag [scalar] (fig2_g2) to [edge label = \text{$p_1$}] (fig2_f1);
	\propag [anti fermion] (fig2_g1) to [edge label = \text{$p_2$}] (fig2_f2);
\end{feynhand}
\end{tikzpicture}
\end{center}
\caption{$s$-channel (left) and $t$-channel (right) diagrams of the Compton-like process $Z'\nu_\alpha \leftrightarrow \phi\bar{\nu}_\beta$. 
Similar diagrams appear in the process $Z' \bar{\nu}_\alpha \leftrightarrow \phi \nu_\beta$.}
\label{fig:diag_Compton}
\end{figure}

Fig.~\ref{fig:diag_Compton} shows the diagrams of the Compton-like scattering $Z' \nu_\alpha \leftrightarrow \phi \bar{\nu}_\beta$ and $Z' \bar{\nu}_\alpha \leftrightarrow \phi \nu_\beta$ ($\alpha$, $\beta$=$\mu$,$\tau$). The right diagram of Fig,~\ref{fig:diag_Compton} corresponds to the $t$-channel process of $Z' \nu_\alpha \leftrightarrow \phi \bar{\nu}_\beta$.

The squared amplitude corresponding to the diagrams is calculated as
\begin{align}
\label{eq:amp_Compton}
   \sum_{\rm spins} & |\mathcal{M}_{Z'\nu_\alpha \leftrightarrow \phi\bar{\nu}_\beta}|^2 \nonumber \\
   &= \frac{16 g_\alpha^2 |h_{\alpha\beta}|^2}
   {(p_3 + p_4)^4+\epsilon^2} 
   \left[4 (p_3 \cdot p_4) (p_3 \cdot p_2) 
   + 4 (p_3 \cdot p_4) (p_4 \cdot p_2) 
   - m_{Z'}^2 (p_4 \cdot p_2) 
   + \frac{4}{m_{Z'}^2} (p_3 \cdot p_4)^2 (p_4 \cdot p_2) \right] \nonumber \\
   & \quad + \frac{16 g_\beta^2 |h_{\alpha\beta}|^2}{(p_3 - p_2)^4+\epsilon^2} 
   \left[4 (p_3 \cdot p_4) (p_3 \cdot p_2) 
   - 4 (p_3 \cdot p_2) (p_4 \cdot p_2) 
   - m_{Z'}^2 (p_4 \cdot p_2) 
   + \dfrac{4}{m_{Z'}^2} (p_3 \cdot p')^2 (p_4 \cdot p_2) \right] \nonumber \\
   & \quad + \frac{16 g_\alpha g_\beta |h_{\alpha\beta}|^2}{(p_3 + p_4)^2(p_3 - p_4)^2+\epsilon^2} 
   \Big[4 (p_3 \cdot p_4) (p_3 \cdot p_2) 
   - 2 (p_3 \cdot p_4) (p_4 \cdot p_2) 
   + 2 (p_3 \cdot p_2) (p_4 \cdot p_2)
   + 4 (p_4 \cdot p_2)^2 \nonumber \\
   & \hspace{140pt} 
   + m_{Z'}^2 (p_4 \cdot p_2) 
   -\frac{4}{m_{Z'}^2}(p_3 \cdot p_4) (p_3 \cdot p_2) (p_4 \cdot p_2) \Big]~,  
\end{align}
where $p_1$, $p_2$, $p_3$, and $p_4$ denote the momentum of the neutrino in the initial states, the neutrino in the final states, U(1)$_{L_{\mu}-L_{\tau}}$ gauge boson and Majoron, respectively. 
The coupling constants $g_\alpha$ depend on the lepton flavor as 
\begin{equation}
g_\alpha = \left\{
\begin{array}{ll}
g_{Z'}^{} & (\alpha = \mu) \\
-g_{Z'}^{} & (\alpha = \tau)  \\
0 & (\textrm{others}) 
\end{array} \right.~.
\end{equation}
In this expression, the factors $(p_3 -p_2)^4$ and $(p_3 -p_2)^2$ in the denominators exhibit the $t$-channel singularity.

The divergence appears in the integration of the momentum, which is needed for evaluating the energy transfer rates. This divergence can be understood physically: the decay of the U(1)$_{L_{\mu}-L_{\tau}}$ gauge boson $Z'\to \nu \bar{\nu}$ generates on-shell neutrinos and causes additional inverse decay of Majoron $\nu \nu \to \phi$. 

Here, we discuss its outcomes in cosmological contexts below. To evaluate the Majoron production efficiency, we solve the Boltzmann equation. 
Its general expression for a particle species $a$ is
\begin{equation}
\frac{\dd n_{a}}{\dd t}=-3 H n_{a}+\frac{\delta n_{a}}{\delta t} \, , 
\end{equation}
where $H$ is the Hubble parameter 
and ${\delta n_{a} / \delta t}$
denotes the change of a particle number density in comoving volume which is generated by collision terms. 
For simplicity, we assume the Maxwell-Boltzmann distribution and neglect the chemical potential and the neutrino mass. 
Collision termss for Majoron production are composed of those from the inverse decay (ID) and the scattering (S).

\begin{equation}
\frac{\delta n_{a}}{\delta t}
=
\left.\frac{\delta n_{a}}{\delta t}\right|_{\rm S}
+
\left.\frac{\delta n_{a}}{\delta t}\right|_{\rm ID} \, , 
\end{equation}
and more concretely,
\begin{align}
    \left.\frac{\delta n_{a}}{\delta t}\right|_{\rm ID} &=-\frac{1}{ S_{i j}} \int \dfrac{\dd ^3\bm{p}_a }{ (2\pi)^32E_a} \int  \dfrac{\dd ^3\bm{p}_i }{ (2\pi)^32E_i} \dfrac{\dd ^3\bm{p}_j }{ (2\pi)^32E_j}(2 \pi)^4 \delta^{(4)}\left(p_a-p_i-p_j\right) \sum_{\text {spins}}\left|\mathcal{M}_{a \leftrightarrow ij}\right|^2 \Lambda_{\rm ID}\{f_i^{\rm MB}\}, 
    \\
	\left.\frac{\delta n_{a}}{\delta t}\right|_{\rm S} &= -\int \dfrac{\dd ^3\bm{p}_a }{ (2\pi)^32E_a}\dfrac{\dd^3\bm{p}_{b} }{(2\pi)^32E_{b}} \dfrac{\dd^3\bm{p}_{c} }{(2\pi)^3 2E_{c}} \dfrac{\dd^3\bm{p}_{d} }{(2\pi)^32E_{d}} 
    \Lambda_S{\{ f_i^{\rm MB}\}}(2\pi)^4\delta ^{(4)}(\bm{p}_{a} + \bm{p}_{b} - \bm{p}_{c} + \bm{p}_{d}  )\sum_{\rm spins} |\mathcal{M}_{ab \leftrightarrow cd}|^2 \, ,
\end{align}
where
\begin{align}
    \Lambda_{\rm ID}\{f_i^{\rm MB}\}&=f^{\rm MB}_a\left(1 \pm f^{\rm MB}_i\right) \left(1 \pm f^{\rm MB}_j\right) -f^{\rm MB}_i f^{\rm MB}_j\left(1 \pm f^{\rm MB}_a\right),
    \\
	\Lambda_{\rm S}\{ f_i^{\rm MB}\} &=  f_a^{\rm MB}f_b^{\rm MB}(1 \pm f_c^{\rm MB})(1 \pm f_d^{\rm MB}) - f_c^{\rm MB}f_d^{\rm MB}(1 \pm f_a^{\rm MB})(1 \pm f_b^{\rm MB}) \, , 
\end{align}
denote the parts corresponding to the distribution function. 
In these expressions with the Maxwell-Boltzmann distribution $f_i^{\rm MB}$, the statistical properties of particles are correctly incorporated by taking $(+)$ and $(-)$ for bosons and fermions, respectively.  
The symmetry factor $S_{ij}$ is equal to two for $i = j$ and one for $i \neq j$.
In the following, $(a,b,c,d)=(\phi,\bar{\nu},Z',\nu)$, or $(1,2,3,4)$.

Writing down Eq.~(23) for Majoron $\phi$ explicitly, we obtain 
\begin{align}
\left.\frac{\delta n_{\phi}}{\delta t}\right|_{\rm S} 
& =-\int \dd \Pi_{1}\dd \Pi_{2} \dd \Pi_{3} \dd \Pi_{4} \Lambda\left(\left\{f_{i}\right\}\right)(2 \pi)^{4} \delta^{(4)}\left(p_{1}+p_{2}-p_{3}-p_{4}\right) \sum_{\text {spins }}|\mathcal{M}|^{2} 
\\ 
&=-\int \dd \Pi_{1}  \dd \Pi_{2}\left(e^{-\left(E_{1} / T_{\phi}+|\bm{p}_{2} / T_{\nu}\right)}-e^{-\left(E_{1}+|\bm{p}_{2}\right) / T_{\nu}}\right) I\, .
\end{align}
where
\begin{align}
\Lambda\left(\left\{f_{i}\right\}\right) &= 
f_{1} f_{2}\left(1 \pm f_{3}\right)\left(1 \pm f_{4}\right)-f_{3} f_{4}\left(1 \pm f_{1}\right)\left(1 \pm f_{2}\right)\, ,
\\
I &= 
\int \dd \Pi_{3} \dd \Pi_{4}(2 \pi)^{4} \delta^{(4)}\left(p_{1}+p_{2}-p_{3}-p_{4}\right) \sum\left|\mathcal{M}_{\nu_{\alpha}  \phi \leftrightarrow Z'\nu_{\beta}}\right|^{2}\, .
\end{align}
By integrating it over $\dd \Pi_{4}$ and $\dd \phi_3$ in the center-of-mass frame, the integral $I$ reduces to
\begin{equation}
I= \left. \frac{1}{2 \pi}\int \dd |\bm{p}_{3} |\dd\cos{\theta_{\rm CM}} \, \dfrac{|\bm{p}_{3}|^2}{2|\bm{p}_4| 2 E_{3}}  \delta\left(\sqrt{s}-E_{3}-|\bm{p}_{4}|\right) \sum\left|\mathcal{M}_{\nu_{\alpha}  \phi \leftrightarrow Z'\nu_{\beta}}\right|^{2}\right|_{\bm{p}_4=-\bm{p}_3}\, .
\end{equation}
The $t$-channel singularity is identified as the second term of the right-hand side of Eq.~\eqref{eq:amp_Compton}, hereafter, we refer to it with $I_t$. 
The expression is 
\begin{align}
I_t(\varepsilon)
&=
\dfrac{16 g_{\beta}^{2}\left|h_{\alpha \beta}\right|^{2}}{2 \pi\sqrt{s}|\bm{p}_2|}  \theta(\sqrt{s}-m_{Z'}) 
\notag
\\
&\quad 
\times \int \dd x \left[
\frac{1}{x^{2}+\varepsilon^2}m_{Z'}^2(s-2m^2_{Z'}+2|\bm{p}_2|\sqrt{s})
-\frac{x}{x^{2}+\varepsilon^2}(s-2m^2_{Z'})+\frac{x^2}{x^{2}+\varepsilon^2}\dfrac{1}{2m_{Z'}^2}(m_{Z'}^2-2|\bm{p}_2|\sqrt{s}-x) \right]
\\
&=
\dfrac{16 g_{\beta}^{2}\left|h_{\alpha \beta}\right|^{2}}{2 \pi\sqrt{s}|\bm{p}_2|}  \theta(\sqrt{s}-m_{Z'}) \times \left[
\dfrac{1}{\varepsilon}\tan^{-1}\dfrac{x}{\varepsilon}m_{Z'}^2(s-2m^2_{Z'}+2|\bm{p}_2|\sqrt{s})
-\dfrac{1}{2}\log(x^2+\varepsilon^2)(s-2m^2_{Z'}) \right.
\notag
\\
&\left. \hspace{90mm} +\left(x-\varepsilon \tan^{-1}\dfrac{x}{\varepsilon}\right)\dfrac{1}{2m_{Z'}^2}(m_{Z'}^2-2|\bm{p}_2|\sqrt{s}-x)\right]~, \label{eq:t-sing}
\end{align}
where the variable $x$ is defined as $x= \left(p_3-p_2\right)^{2}$.
The integration is performed from $m_{Z'}^2-2|\bm{p}_2|\sqrt{s} $ to $ m_{Z'}^2-2|\bm{p}_2|\dfrac{m_{Z^{\prime}}^{2}}{\sqrt{s}}$. 
The first term of Eq.~(\ref{eq:t-sing}) is obtained by applying the scheme in Sec.~\ref{subsec:analytic-prescription}to the first term of Eq.~(\ref{eq:t-sing}). 
In the second and third terms, no divergence appears in the limit $\varepsilon \to 0$, hence no special cares are needed. 
We define $I_0$ as $I_t$ without the divergent part:
\begin{align}
&I_0
=
\dfrac{16 g_{\beta}^{2}\left|h_{\alpha \beta}\right|^{2}}{2 \pi\sqrt{s}|\bm{p}_2|}  \theta(\sqrt{s}-m_{Z'}) 
\left[
-\dfrac{1}{x}m_{Z'}^2(s-2m^2_{Z'}+2|\bm{p}_2|\sqrt{s})
-\dfrac{1}{2}\log(x^2+\varepsilon^2)(s-2m^2_{Z'}) \right. \nonumber \\
& \left. \hspace{110mm} +\dfrac{x}{2m_{Z'}^2}(m_{Z'}^2-2|\bm{p}_2|\sqrt{s}-x)\right]\, ,
\end{align}
and this is what we need for evaluating the production efficiency.

For Majoron $\phi$, the interaction rates in the thermal equilibrium expressed as $\langle \Gamma_{\rm S} \rangle$, $\langle \Gamma_{\rm ID} \rangle$ are
\begin{equation}
\langle \Gamma_{\rm S} \rangle = \frac{1}{n_{\phi,\rm eq}} \left.\frac{\delta n_{\phi}}{\delta t}\right|_{\rm S} 
, \quad 
\langle \Gamma_{\rm ID} \rangle = 
\frac{1}{n_{\phi,\rm eq}} 
\left.\frac{\delta n_{\phi}}{\delta t}\right|_{\rm ID} ,
\label{eq:gamma}
\end{equation}
where $n_{\phi,{\rm eq}}$ is 
the number density of the species $\phi$ in the equilibrium. By applying the Gamow's criteria to these quantities, the decoupling of the interactions are evaluated. 
The process is in the thermal equilibrium for $\langle \Gamma_{{\rm S}, {\rm ID}} \rangle / H > 1$. 
Here, we assume that the Majoron $\phi$ did not exist in the early Universe and consider the net contribution from Majoron production direction.The Hubble parameter is calculated using the effective degree of freedom taken from Ref. \cite{Saikawa:2020swg}.

\begin{figure}[tbp]
\begin{center}
    \includegraphics[width=10cm]{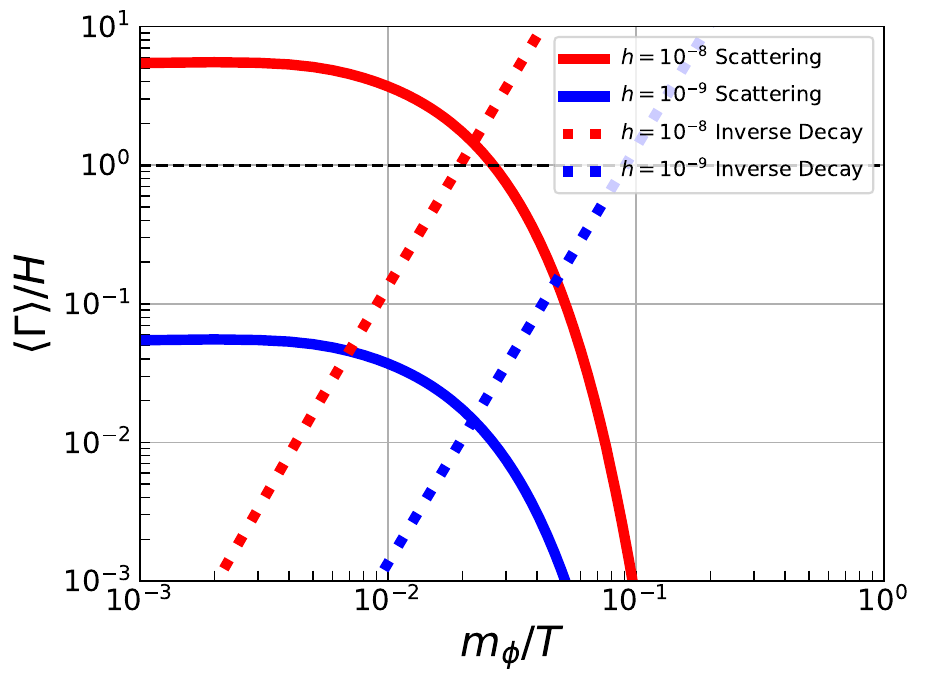}
\end{center}
\caption{Comparison between the production efficiencies of the inverse decay and the scattering. Horizontal axis is the temperature of the Universe normalized by the Majoron mass, and the vertical axis shows the interaction rate. The scattering rate can dominate over the inverse decay rate at high temperatures. We fix the Majoron mass $m_\phi=0.1$~MeV. 
We limit our discussion to the regime after the QCD phase transition.}
\label{fig:result}
\end{figure}

Figure~\ref{fig:result} compares the efficiency of the scattering and the inverse decay processes. 
The vertical axis represents the ratio of the interaction rates $\langle \Gamma_{\rm S, ID} \rangle$ to the Hubble parameter $H$, and the horizontal axis is the temperature evolution normalized by the Majoron mass. Here, all of the Majoron-neutrino couplings $h_{\alpha \beta}$ in Eq.~\eqref{eq:Lag-phi} are assumed to be the same for simplicity.
The masses of $Z'$ and $\phi$ are fixed to $m_{Z'} = 13$~MeV  and $m_\phi = 0.1$~MeV in this figure. The Gamow's criteria is shown as the dotted(black) horizontal line. Above the dotted line, the interaction is in thermal equilibrium while below the line it is regarded as being decoupled. 
As shown in the Fig.~\ref{fig:result}, the scattering can dominate over the inverse decay at high temperatures, and it can satisfy the Gamow's criteria for $h=10^{-8}$, i.e., the process is efficient enough for the Majoron production. 
In other words, the picture neglecting the scattering contributions for the Majoron production can be applicable only for $h<10^{-9}$. 
This bound is stronger than those obtained in previous works~\cite{Escudero:2019gvw, EscuderoAbenza:2020cmq}.

\section{Summary}
\label{s:discussion}

In this work, we propose an analytic method to remove the divergence in the $t$-channel singularity without introducing any new quantities such as temperature cutoffs or beam widths. 
Such a divergence appears when the intermediate stable particle becomes an on-shell state. 
The expansion of the integrated scattering amplitude in terms of the Feynman prescription parameter $\epsilon$ yields terms proportional to $1/\epsilon$, which is identified as the contribution from the real particle. 
By removing these terms, we can successfully separate the contributions from the scattering and the inverse decay or decays in particle production.

As an demonstration, we apply the scheme to the U(1)$_{L_\mu - L_\tau} \times$ U(1)$_L$ model. 
In previous works, the Majoron has been considered to be solely produced by the inverse decay. 
However, by separating the contribution from the scattering processes, 
we find that the scattering can be frequent enough and dominate over the inverse decay for producing Majorons in the early Universe. 
Its phenomenological implications to the U(1)$_{L_\mu - L_\tau} \times$ U(1)$_L$ models are discussed in our future works. 
At high temperatures, incorporating finite-temperature effects within quantum field theory is essential to achieve quantitative agreement with previous studies~\cite{Grzadkowski:2021kgi,Iglicki:2022jjf}. 
In contrast, at low temperatures, corrections from particles such as neutrinos are expected to be negligible. 
To emphasize the broad applicability of the analytic framework developed in this work across a wide range of contexts, we defer a detailed quantitative comparison at high temperatures to future studies.
The scheme of this work can also be applied in much wider contexts, such as giving an interpretation of the finite-beam size effects in muon collider experiments with mathematical support.

\section*{Acknowledgements}
We gratefully thank Makiko Nio for discussions at an earlier stage of this work. 
This work was partially supported by JSPS KAKENHI Grant Number JP23K13097 (K.A.), JP25KJ0401 (K.A.), JP20H05852(N.H.), JP22K14035(N.H.) and JP25H01524(J.S.). 
The works of N.H. and M.J.S.Y. are partly supported by the MEXT Leading Initiative for Excellent Young Researchers Grant Number JP2023L0013.

\section*{DATA AVAILABILITY}
Plot data is available upon request.

\bibliographystyle{apsrev}
\bibliography{ref}

\begin{thebibliography}{28}
\expandafter\ifx\csname natexlab\endcsname\relax\def\natexlab#1{#1}\fi
\expandafter\ifx\csname bibnamefont\endcsname\relax
  \def\bibnamefont#1{#1}\fi
\expandafter\ifx\csname bibfnamefont\endcsname\relax
  \def\bibfnamefont#1{#1}\fi
\expandafter\ifx\csname citenamefont\endcsname\relax
  \def\citenamefont#1{#1}\fi
\expandafter\ifx\csname url\endcsname\relax
  \def\url#1{\texttt{#1}}\fi
\expandafter\ifx\csname urlprefix\endcsname\relax\def\urlprefix{URL }\fi
\providecommand{\bibinfo}[2]{#2}
\providecommand{\eprint}[2][]{\url{#2}}

\bibitem[{\citenamefont{Lehmann et~al.}(1955)\citenamefont{Lehmann, Symanzik,
  and Zimmermann}}]{Lehmann:1954rq}
\bibinfo{author}{\bibfnamefont{H.}~\bibnamefont{Lehmann}},
  \bibinfo{author}{\bibfnamefont{K.}~\bibnamefont{Symanzik}}, \bibnamefont{and}
  \bibinfo{author}{\bibfnamefont{W.}~\bibnamefont{Zimmermann}},
  \bibinfo{journal}{Nuovo Cim.} \textbf{\bibinfo{volume}{1}},
  \bibinfo{pages}{205} (\bibinfo{year}{1955}).

\bibitem[{\citenamefont{Cutkosky}(1960)}]{Cutkosky:1960sp}
\bibinfo{author}{\bibfnamefont{R.~E.} \bibnamefont{Cutkosky}},
  \bibinfo{journal}{J. Math. Phys.} \textbf{\bibinfo{volume}{1}},
  \bibinfo{pages}{429} (\bibinfo{year}{1960}).

\bibitem[{\citenamefont{Mandelstam}(1958)}]{Mandelstam:1958xc}
\bibinfo{author}{\bibfnamefont{S.}~\bibnamefont{Mandelstam}},
  \bibinfo{journal}{Phys. Rev.} \textbf{\bibinfo{volume}{112}},
  \bibinfo{pages}{1344} (\bibinfo{year}{1958}).

\bibitem[{\citenamefont{Breit and Wigner}(1936)}]{Breit:1936zzb}
\bibinfo{author}{\bibfnamefont{G.}~\bibnamefont{Breit}} \bibnamefont{and}
  \bibinfo{author}{\bibfnamefont{E.}~\bibnamefont{Wigner}},
  \bibinfo{journal}{Phys. Rev.} \textbf{\bibinfo{volume}{49}},
  \bibinfo{pages}{519} (\bibinfo{year}{1936}).

\bibitem[{\citenamefont{Yoshimura}(1978)}]{Yoshimura:1978ex}
\bibinfo{author}{\bibfnamefont{M.}~\bibnamefont{Yoshimura}},
  \bibinfo{journal}{Phys. Rev. Lett.} \textbf{\bibinfo{volume}{41}},
  \bibinfo{pages}{281} (\bibinfo{year}{1978}), \bibinfo{note}{[Erratum:
  Phys.Rev.Lett. 42, 746 (1979)]}.

\bibitem[{\citenamefont{Fukugita and Yanagida}(1986)}]{Fukugita:1986hr}
\bibinfo{author}{\bibfnamefont{M.}~\bibnamefont{Fukugita}} \bibnamefont{and}
  \bibinfo{author}{\bibfnamefont{T.}~\bibnamefont{Yanagida}},
  \bibinfo{journal}{Phys. Lett. B} \textbf{\bibinfo{volume}{174}},
  \bibinfo{pages}{45} (\bibinfo{year}{1986}).

\bibitem[{\citenamefont{Kolb and Wolfram}(1980)}]{Kolb:1979qa}
\bibinfo{author}{\bibfnamefont{E.~W.} \bibnamefont{Kolb}} \bibnamefont{and}
  \bibinfo{author}{\bibfnamefont{S.}~\bibnamefont{Wolfram}},
  \bibinfo{journal}{Nucl. Phys. B} \textbf{\bibinfo{volume}{172}},
  \bibinfo{pages}{224} (\bibinfo{year}{1980}), \bibinfo{note}{[Erratum:
  Nucl.Phys.B 195, 542 (1982)]}.

\bibitem[{\citenamefont{Giudice et~al.}(2004)\citenamefont{Giudice, Notari,
  Raidal, Riotto, and Strumia}}]{Giudice:2003jh}
\bibinfo{author}{\bibfnamefont{G.~F.} \bibnamefont{Giudice}},
  \bibinfo{author}{\bibfnamefont{A.}~\bibnamefont{Notari}},
  \bibinfo{author}{\bibfnamefont{M.}~\bibnamefont{Raidal}},
  \bibinfo{author}{\bibfnamefont{A.}~\bibnamefont{Riotto}}, \bibnamefont{and}
  \bibinfo{author}{\bibfnamefont{A.}~\bibnamefont{Strumia}},
  \bibinfo{journal}{Nucl. Phys. B} \textbf{\bibinfo{volume}{685}},
  \bibinfo{pages}{89} (\bibinfo{year}{2004}), \eprint{hep-ph/0310123}.

\bibitem[{\citenamefont{Peierls}(1961)}]{Peierls:1961zz}
\bibinfo{author}{\bibfnamefont{R.~F.} \bibnamefont{Peierls}},
  \bibinfo{journal}{Phys. Rev. Lett.} \textbf{\bibinfo{volume}{6}},
  \bibinfo{pages}{641} (\bibinfo{year}{1961}).

\bibitem[{\citenamefont{Brayshaw et~al.}(1978)\citenamefont{Brayshaw, Simmons,
  and Tuan}}]{Brayshaw:1978xt}
\bibinfo{author}{\bibfnamefont{D.~D.} \bibnamefont{Brayshaw}},
  \bibinfo{author}{\bibfnamefont{W.~A.} \bibnamefont{Simmons}},
  \bibnamefont{and} \bibinfo{author}{\bibfnamefont{S.~F.} \bibnamefont{Tuan}},
  \bibinfo{journal}{Phys. Rev. D} \textbf{\bibinfo{volume}{18}},
  \bibinfo{pages}{1719} (\bibinfo{year}{1978}).

\bibitem[{\citenamefont{Melnikov and Serbo}(1996)}]{Melnikov:1996na}
\bibinfo{author}{\bibfnamefont{K.}~\bibnamefont{Melnikov}} \bibnamefont{and}
  \bibinfo{author}{\bibfnamefont{V.~G.} \bibnamefont{Serbo}},
  \bibinfo{journal}{Phys. Rev. Lett.} \textbf{\bibinfo{volume}{76}},
  \bibinfo{pages}{3263} (\bibinfo{year}{1996}), \eprint{hep-ph/9601221}.

\bibitem[{\citenamefont{Melnikov and Serbo}(1997)}]{Melnikov:1996iu}
\bibinfo{author}{\bibfnamefont{K.}~\bibnamefont{Melnikov}} \bibnamefont{and}
  \bibinfo{author}{\bibfnamefont{V.~G.} \bibnamefont{Serbo}},
  \bibinfo{journal}{Nucl. Phys. B} \textbf{\bibinfo{volume}{483}},
  \bibinfo{pages}{67} (\bibinfo{year}{1997}), \bibinfo{note}{[Erratum:
  Nucl.Phys.B 662, 409 (2003)]}, \eprint{hep-ph/9601290}.

\bibitem[{\citenamefont{Melnikov et~al.}(1996)\citenamefont{Melnikov, Kotkin,
  and Serbo}}]{Melnikov:1996ft}
\bibinfo{author}{\bibfnamefont{K.}~\bibnamefont{Melnikov}},
  \bibinfo{author}{\bibfnamefont{G.~L.} \bibnamefont{Kotkin}},
  \bibnamefont{and} \bibinfo{author}{\bibfnamefont{V.~G.} \bibnamefont{Serbo}},
  \bibinfo{journal}{Phys. Rev. D} \textbf{\bibinfo{volume}{54}},
  \bibinfo{pages}{3289} (\bibinfo{year}{1996}), \eprint{hep-ph/9603352}.

\bibitem[{\citenamefont{Dams and Kleiss}(2003)}]{Dams:2002uy}
\bibinfo{author}{\bibfnamefont{C.}~\bibnamefont{Dams}} \bibnamefont{and}
  \bibinfo{author}{\bibfnamefont{R.}~\bibnamefont{Kleiss}},
  \bibinfo{journal}{Eur. Phys. J. C} \textbf{\bibinfo{volume}{29}},
  \bibinfo{pages}{11} (\bibinfo{year}{2003}), \eprint{hep-ph/0212301}.

\bibitem[{\citenamefont{Dams and Kleiss}(2004)}]{Dams:2003gn}
\bibinfo{author}{\bibfnamefont{C.}~\bibnamefont{Dams}} \bibnamefont{and}
  \bibinfo{author}{\bibfnamefont{R.}~\bibnamefont{Kleiss}},
  \bibinfo{journal}{Eur. Phys. J. C} \textbf{\bibinfo{volume}{36}},
  \bibinfo{pages}{177} (\bibinfo{year}{2004}), \eprint{hep-ph/0309336}.

\bibitem[{\citenamefont{Nowakowski and Pilaftsis}(1993)}]{Nowakowski:1993iu}
\bibinfo{author}{\bibfnamefont{M.}~\bibnamefont{Nowakowski}} \bibnamefont{and}
  \bibinfo{author}{\bibfnamefont{A.}~\bibnamefont{Pilaftsis}},
  \bibinfo{journal}{Z. Phys. C} \textbf{\bibinfo{volume}{60}},
  \bibinfo{pages}{121} (\bibinfo{year}{1993}), \eprint{hep-ph/9305321}.

\bibitem[{\citenamefont{Karamitros and Pilaftsis}(2023)}]{Karamitros:2022nnh}
\bibinfo{author}{\bibfnamefont{D.}~\bibnamefont{Karamitros}} \bibnamefont{and}
  \bibinfo{author}{\bibfnamefont{A.}~\bibnamefont{Pilaftsis}},
  \bibinfo{journal}{Phys. Rev. D} \textbf{\bibinfo{volume}{108}},
  \bibinfo{pages}{036007} (\bibinfo{year}{2023}), \eprint{2208.10425}.

\bibitem[{\citenamefont{Grzadkowski et~al.}(2022)\citenamefont{Grzadkowski,
  Iglicki, and Mr\'owczy\'nski}}]{Grzadkowski:2021kgi}
\bibinfo{author}{\bibfnamefont{B.}~\bibnamefont{Grzadkowski}},
  \bibinfo{author}{\bibfnamefont{M.}~\bibnamefont{Iglicki}}, \bibnamefont{and}
  \bibinfo{author}{\bibfnamefont{S.}~\bibnamefont{Mr\'owczy\'nski}},
  \bibinfo{journal}{Nucl. Phys. B} \textbf{\bibinfo{volume}{984}},
  \bibinfo{pages}{115967} (\bibinfo{year}{2022}), \eprint{2108.01757}.

\bibitem[{\citenamefont{Iglicki}(2023)}]{Iglicki:2022jjf}
\bibinfo{author}{\bibfnamefont{M.}~\bibnamefont{Iglicki}},
  \bibinfo{journal}{JHEP} \textbf{\bibinfo{volume}{06}}, \bibinfo{pages}{006}
  (\bibinfo{year}{2023}), \eprint{2212.00561}.

\bibitem[{\citenamefont{Ginzburg}(1996)}]{Ginzburg:1995bc}
\bibinfo{author}{\bibfnamefont{I.~F.} \bibnamefont{Ginzburg}},
  \bibinfo{journal}{Nucl. Phys. B Proc. Suppl.} \textbf{\bibinfo{volume}{51}},
  \bibinfo{pages}{85} (\bibinfo{year}{1996}), \eprint{hep-ph/9601272}.

\bibitem[{\citenamefont{Escudero et~al.}(2019)\citenamefont{Escudero, Hooper,
  Krnjaic, and Pierre}}]{Escudero:2019gzq}
\bibinfo{author}{\bibfnamefont{M.}~\bibnamefont{Escudero}},
  \bibinfo{author}{\bibfnamefont{D.}~\bibnamefont{Hooper}},
  \bibinfo{author}{\bibfnamefont{G.}~\bibnamefont{Krnjaic}}, \bibnamefont{and}
  \bibinfo{author}{\bibfnamefont{M.}~\bibnamefont{Pierre}},
  \bibinfo{journal}{JHEP} \textbf{\bibinfo{volume}{03}}, \bibinfo{pages}{071}
  (\bibinfo{year}{2019}), \eprint{1901.02010}.

\bibitem[{\citenamefont{Escudero}(2019)}]{Escudero:2018mvt}
\bibinfo{author}{\bibfnamefont{M.}~\bibnamefont{Escudero}},
  \bibinfo{journal}{JCAP} \textbf{\bibinfo{volume}{02}}, \bibinfo{pages}{007}
  (\bibinfo{year}{2019}), \eprint{1812.05605}.

\bibitem[{\citenamefont{Asai et~al.}(2024)\citenamefont{Asai, Asano, Sato, and
  Yang}}]{Asai:2023ajh}
\bibinfo{author}{\bibfnamefont{K.}~\bibnamefont{Asai}},
  \bibinfo{author}{\bibfnamefont{T.}~\bibnamefont{Asano}},
  \bibinfo{author}{\bibfnamefont{J.}~\bibnamefont{Sato}}, \bibnamefont{and}
  \bibinfo{author}{\bibfnamefont{M.~J.~S.} \bibnamefont{Yang}},
  \bibinfo{journal}{PTEP} \textbf{\bibinfo{volume}{2024}},
  \bibinfo{pages}{073E01} (\bibinfo{year}{2024}), \eprint{2309.01162}.

\bibitem[{\citenamefont{Araki et~al.}(2019)\citenamefont{Araki, Asai, Sato, and
  Shimomura}}]{Araki:2019rmw}
\bibinfo{author}{\bibfnamefont{T.}~\bibnamefont{Araki}},
  \bibinfo{author}{\bibfnamefont{K.}~\bibnamefont{Asai}},
  \bibinfo{author}{\bibfnamefont{J.}~\bibnamefont{Sato}}, \bibnamefont{and}
  \bibinfo{author}{\bibfnamefont{T.}~\bibnamefont{Shimomura}},
  \bibinfo{journal}{Phys. Rev. D} \textbf{\bibinfo{volume}{100}},
  \bibinfo{pages}{095012} (\bibinfo{year}{2019}), \eprint{1909.08827}.

\bibitem[{\citenamefont{Araki et~al.}(2021)\citenamefont{Araki, Asai, Honda,
  Kasuya, Sato, Shimomura, and Yang}}]{Araki:2021xdk}
\bibinfo{author}{\bibfnamefont{T.}~\bibnamefont{Araki}},
  \bibinfo{author}{\bibfnamefont{K.}~\bibnamefont{Asai}},
  \bibinfo{author}{\bibfnamefont{K.}~\bibnamefont{Honda}},
  \bibinfo{author}{\bibfnamefont{R.}~\bibnamefont{Kasuya}},
  \bibinfo{author}{\bibfnamefont{J.}~\bibnamefont{Sato}},
  \bibinfo{author}{\bibfnamefont{T.}~\bibnamefont{Shimomura}},
  \bibnamefont{and} \bibinfo{author}{\bibfnamefont{M.~J.~S.}
  \bibnamefont{Yang}}, \bibinfo{journal}{PTEP} \textbf{\bibinfo{volume}{2021}},
  \bibinfo{pages}{103B05} (\bibinfo{year}{2021}), \eprint{2103.07167}.

\bibitem[{\citenamefont{Saikawa and Shirai}(2020)}]{Saikawa:2020swg}
\bibinfo{author}{\bibfnamefont{K.}~\bibnamefont{Saikawa}} \bibnamefont{and}
  \bibinfo{author}{\bibfnamefont{S.}~\bibnamefont{Shirai}},
  \bibinfo{journal}{JCAP} \textbf{\bibinfo{volume}{08}}, \bibinfo{pages}{011}
  (\bibinfo{year}{2020}), \eprint{2005.03544}.

\bibitem[{\citenamefont{Escudero and Witte}(2020)}]{Escudero:2019gvw}
\bibinfo{author}{\bibfnamefont{M.}~\bibnamefont{Escudero}} \bibnamefont{and}
  \bibinfo{author}{\bibfnamefont{S.~J.} \bibnamefont{Witte}},
  \bibinfo{journal}{Eur. Phys. J. C} \textbf{\bibinfo{volume}{80}},
  \bibinfo{pages}{294} (\bibinfo{year}{2020}), \eprint{1909.04044}.

\bibitem[{\citenamefont{Escudero~Abenza}(2020)}]{EscuderoAbenza:2020cmq}
\bibinfo{author}{\bibfnamefont{M.}~\bibnamefont{Escudero~Abenza}},
  \bibinfo{journal}{JCAP} \textbf{\bibinfo{volume}{05}}, \bibinfo{pages}{048}
  (\bibinfo{year}{2020}), \eprint{2001.04466}.

\end{thebibliography}

\end{document}